\documentclass[superscriptaddress,groupedaddress,nofootnoteinbib,12pt]{article} 
\pdfoutput=1
\usepackage{graphicx}
\usepackage{color}
\usepackage{dcolumn}   
\usepackage{bm}       
\usepackage{amssymb}  
\usepackage{amsmath}
\usepackage{sectsty}
\usepackage{latexsym}
\usepackage{float}
\usepackage{ifthen}
\usepackage{caption,subfig}
\usepackage{enumerate}
\usepackage{url}
\usepackage{caption,subfig}
\usepackage{feynmf}	
\usepackage{jcappub}
\usepackage{amsopn}
\usepackage{float}
\usepackage{pst-pdf}
\usepackage{amsfonts}
\usepackage{multirow}
\usepackage{array}
\usepackage{booktabs}
\usepackage{rotating}

\def\clap#1{\hbox to 0pt{\hss#1\hss}}

\def\({\left(}
\def\){\right)}
\def\[{\left[}
\def\]{\right]}
\def\bea{\begin{eqnarray}}
\def\eea{\end{eqnarray}}
\def\be{\begin{equation}}
\def\ee{\end{equation}}
\def\ba{\begin{eqnarray}}
\def\ea{\end{eqnarray}}
\def\beq{\begin{eqnarray}}
\def\eeq{\end{eqnarray}}

\def\d{\mathrm{d}}

\def\clap#1{\hbox to 0pt{\hss#1\hss}}

\definecolor{forestgreen}{rgb}{0.133,0.545,0.133}
\newboolean{editorial}
\setboolean{editorial}{true}
\newcommand{\editorial}[2]{\ifthenelse{\boolean{editorial}}{\textcolor{red}{[\textsf{\textbf{{#1}}}: }\textcolor{blue}{\textsf{{#2}}}\textcolor{red}{]}}{}}

\renewcommand{\d}{\mathrm{d}}

 \def\be   {\begin{equation}}   \def\ee   {\end{equation}}

 \def\ba  {\begin{eqnarray}}   \def\ea  {\end{eqnarray}}

\renewcommand{\thesubsection}{\arabic{section}.\arabic{subsection}}

\begin{document}

\title{Quantum Corrections in Galileons from Matter Loops}

\author{Lavinia Heisenberg,$^{a,b}$}
\affiliation{$^{a}$Perimeter Institute for Theoretical Physics, \\
31 Caroline St. N, Waterloo, Ontario, Canada, N2L 2Y5}
\affiliation{$^{b}$D\'epartment de Physique  Th\'eorique and Center for Astroparticle Physics,\\
Universit\'e de Gen\`eve, 24 Quai E. Ansermet, CH-1211  Gen\`eve, Switzerland}

	\emailAdd{Lavinia.Heisenberg@unige.ch}

\abstract{
Galileon interactions represent  a class of effective field theories that have received much attention since their inception. They can be treated in their own right as scalar field theories with a specific global shift and Galilean symmetry or as a descendant of a more fundamental theory like massive gravity. It is well known that the Galileon theories are stable under quantum corrections thanks to the non-renormalization theorem which is not due to the symmetry. We consider different covariant couplings of this Galileon scalar field with the matter field: the conformal coupling, the disformal coupling and the longitudinal coupling. We compute the one-loop quantum corrections to the Galileon interactions from the coupling to the external matter fields. In all the considered cases of covariant couplings we show that the terms generated by one-loop matter corrections not only renormalize the Galileon interactions but also give rise to higher order derivative ghost interactions. However, the renormalized version of the Galileon interactions as well as the new interactions come at a scale suppressed by the original classical coupling scale and hence are harmless within the regime of validity of the effective field theory.}


\maketitle


\section{Introduction}
Cosmology has progressively developed from a philosophical to an empirical scientific discipline. Given the high precision achieved by the cosmological observations, cosmology is now a suitable arena to test fundamental physics. It is witnessing promising attempts to unite the physics of the large scale structures in the Universe with the physics of the small scale. Latest observations from Planck \cite{Ade:2013uln} and BICEP2 \cite{Ade:2014xna} with the exquisite precision have driven the models of early universe into a corner. Inflation, which describes a phase of accelerated expansion in the very early Universe might explain the initial conditions of the Universe.  \\

Observations of the Cosmic Microwave Background, supernovae Ia,  Baryon Acoustic Oscillations and lensing have led to the cosmological standard model which also requires an accelerated expansion of the late Universe, driven by dark energy \cite{Hinshaw:2012aka, Percival:2009xn, 0004-637X-690-2-1236, Ade:2013zuv}. The physical origin of the accelerated expansion is still a mystery. In the Standard Model of particles the detection of the missing fundamental particle, the Higgs boson, was a revolutionary event. In a similar way, the missing particles in the Standard Model of cosmology, like the graviton, and the resolution to the puzzle of accelerated expansion and its origin would be as revolutionary. There are several promising explanatory attempts to provide an explanation for the accelerated expansion of the late Universe.\\

One promising approach corresponds to explaining the acceleration of the Universe by modifying the geometrical part of Einstein's equations. Particularly, weakening gravity on cosmological scales might not only be responsible for a late-time speed-up of the Hubble expansion, but could also tackle the cosmological constant problem, which reflects the large discrepancy between observations and theoretical predictions. Such promising scenarios arise in infrared modifications of general relativity like massive gravity or in higher-dimensional frameworks \cite{Dvali:2000hr, deRham:2010ik, deRham:2010kj, deRham:2010tw}. All these models of infrared modifications of general relativity are united by the common feature of invoking new degrees of freedom. These degrees of freedom in the considered space-time are particles characterized by their masses and spins (or equivalently helicities). These particles are the excited quanta of the underlying fields. Their Lagrangian are constructed based on the requirement of yielding second order equations of motion, hence with a bounded Hamiltonian from below. With these requirements another successful and interesting class of infrared modifications was introduced: the Galileons \cite{Nicolis:2008in}. This Galileon model relies strongly on the symmetry of the new scalar degree of freedom, namely the invariance under internal Galilean and shift transformations $\phi\to\phi+b_\mu x^\mu+c$, and on the ghost absence. Interestingly Galileon interactions naturally arise in theories of massive gravity which has also been constructed to be ghost-free \cite{deRham:2010ik, deRham:2010tw, deRham:2014zqa}. The Galileon theory has also been generalized to the non-flat background case. Direct covariantization might give rise to ghost-like terms in the equation of motion, which has motivated the introduction of non-minimal coupling between $\phi$ and the curvature \cite{Deffayet:2009mn, Deffayet:2009wt, deRham:2010eu}. Even if this covariantization is ghost-free, the Galileon symmetry is lost in curved backgrounds. However, there has been also successful generalizations to the maximally symmetric backgrounds with generalized Galileon symmetry \cite{Burrage:2011bt, Goon:2011qf}. Interestingly enough, there has been a recent construction of a covariant Galileon theory without the requirement of non-minimal interactions and still avoiding ghost instabilities \cite{Gleyzes:2014dya,Gleyzes:2014qga, Gao:2014soa,Fasiello:2014aqa}. Another interesting point worth to mention is the fact that the covariant Galileon interactions also arise from the covariantization of the decoupling limit of massive gravity \cite{deRham:2011by, Heisenberg:2014kea}. Similarly, there has been attempts to find the  vector Galileons besides the Maxwell kinetic term with second order equations of motion \cite{Deffayet:2013tca, Heisenberg:2014rta, Tasinato:2014eka}. \\

The Galileon theory exhibits a broad and interesting phenomenology. However one potentially worrying phenomenon is the fact that fluctuations of the Galileon field can propagate superluminally in the regime of interest \cite{Nicolis:2008in, Goon:2010xh, deFromont:2013iwa}, i.e. faster than light. Nevertheless, it has been shown in \cite{Burrage:2011cr} that closed time like curves can never arise since the Galileon inevitably becomes infinitely strongly coupled and breaks down the effective field theory. Additionally, a dual description to the Galileon interactions has been discovered by \cite{deRham:2013hsa, deRham:2014lqa, Kampf:2014rka} in which the original Galileon theory gets mapped to another Galileon theory by a non trivial field redefinition $\tilde x_\mu=x_\mu+\partial_\mu\phi(x)$. For a given very specific Galileon coefficients the Galileon interactions are dual to a free massless scalar field. Thus, this mapping between a free luminal theory and the superluminal Galileon theory suggests that the naive existence of superluminal propagation can still give rise to causal theory with analytic and unitary S-matrix \cite{deRham:2013hsa}. These properties together with the presence of a Vainshtein mechanism could be tied to theories which allow for an alternative to UV completion such as classicalization.\\

Galileon interactions can be considered as an effective field theory constructed by the above mentioned restrictions of symmetry and ghost absence. In order for the theory to be viable, the Vainshtein mechanism is needed, which on the other hand relies on the presence of interactions at an energy scale $\Lambda_3\ll M_{\rm Pl}$. From a traditional effective field theory point of view these interactions are irrelevant operators which renders the theory non-renormalizable, but to contrary to the traditional case, within the Galileon theory these irrelevant operators need to be large in the regime of interest, in the so called strong coupled regime $\partial^2\phi\sim \Lambda_3^3$. Therefore, one might have concerns that the effective field theory could go out of control in this strong coupling regime where the irrelevant operators need to be large. Nevertheless, the Galileon theories are not typical effective field theories in the sense that it is organized in the small parameter expansion of the whole operator but rather it has to be reorganized in a way that the derivative now plays the role of the small parameter rather than the whole operator itself \cite{Nicolis:2008in, Nicolis:2004qq, deRham:2012ew, deRham:2014wfa}. There exist a regime of interest for which $\phi \sim \Lambda_3$, $\partial\phi\sim \Lambda_3^2$ and $\partial^2\phi\sim \Lambda_3^3$ even though any further derivative is suppressed $\partial^3\phi \ll \Lambda_3^4$, meaning that the effective field expansion is reorganized such that the Galileon interactions are the relevant operators with equations of motion with only two derivatives, while all other interactions with equations of motion with more than two derivatives are treated as negligibale corrections. For a similar discussions see also \cite{deRham:2014wfa,deRham:2014naa, deRham:2014fha}.\\

Galileon interactions are protected against quantum corrections via the non-renormalization theorem. First of all, the Galileon and shift symmetry will prevent to generate local operators by Galileon loop corrections which would explicitly break these symmetries, like potential interactions. But this is not enough for the non-renormalization theorem. Quantum corrections might still generate local operators which are invariant under shift and Galileon transformations, either renormalizing Galileon interactions themselves and giving rise to large quantum corrections of the strong coupling scale $\Lambda_3$ or generating operators of higher derivative interactions. It is the non-renormalization theorem, which ensures that the Galileon interactions themselves are not renormalized at all and that the higher derivative operators are irrelevant corrections in the regime of validity of the effective field theory.
The specific form of the counter terms arising in the 1-loop effective action coming from Galileon loops is such that they all come with at least one extra derivative as compared to the original interactions. Therefore there is no counter term which takes the Galileon form, and the Galileon interactions are hence not renormalized. This would mean that the Galileon coupling constants may be technically natural tuned to any value and remain radiatively stable.\\

In this work we address the question of one loop quantum corrections coming from matter loops. We will consider the Galileon scalar field as a scalar field in its own right, without restricting it to the massive gravity case. The Galileon scalar field can couple to matter as a conformal mode, $\phi T$ at the linear level but also as a longitudinal mode, $\partial_\mu \partial_\nu \phi T^{\mu\nu}$ (even though this coupling would vanish for a conserved source), where $T_{\mu\nu}$ is the stress-energy tensor of the external matter field and $T$ is the trace of it. At the non-linear level one could also consider more generic conformal couplings like $\phi^2 T$ or $f(\phi)T$ even though the symmetry would be broken at the level of the equations of motion. Naturally, one could also consider derivative couplings of the form $\partial_\mu \phi \partial_\nu\phi T^{\mu\nu}$ known as disformal coupling (which also arises in massive gravity), or more generally $f(\phi)\partial_\mu \phi \partial_\nu \phi T^{\mu\nu}$. We will be mainly concentrating on the cases of conformal, disformal and longitudinal couplings.


\section{Setup}
\label{sec:setup}
In this section we will set up our framework and notations. For convenience, we will work in Euclidean space such that the Galileon interactions live on top of a flat Euclidean metric $\delta_{ab}$. Furthermore, we will use units for which $\bar h=1$ and we will use the $(-,+,+,+)$ signature convention. For simplicity, we will assume a massive scalar field for our matter field which covariantly couples to the Galileon field. \\
Our starting point is the action for the Galileon $\phi$ and a massive scalar field $\chi$,
\begin{eqnarray}
S&=&\int \d ^4 x \left(\mathcal{L}_{Gal} +\mathcal{L}_{\rm matter}\right)\\
&=& \int \d ^4 x \mathcal{L}_{Gal}+\int \d ^4 x \left(\frac 12 \left(\partial \chi\right)^2+\frac 12 M^2 \chi^2\right)\,.
\end{eqnarray}
where $\mathcal{L}_{Gal}$ is the Lagrangian for the Galileon interactions
\begin{equation}\label{Galileons}
\mathcal{L}_{Gal}=\sum_{n=0}^4 c_n \phi \;\; \mathcal{U}_n(\Phi(x))
\end{equation}
with $\Phi_{\mu\nu}=\partial_\mu\partial_\nu \phi$ and the characteristic symmetric polynomial invariants
\begin{equation}
\mathcal{U}[\Phi]=\mathcal{E}^{\mu_1\dots\mu_4}\mathcal{E}^{\nu_1\dots\nu_4}\prod_{j=1}^n \Phi_{\mu_j \nu_j}\prod_{k=n+1}^4 \eta_{\mu_k \nu_k}
\end{equation}
In order to have the standard kinetic term for the Galileon scalar field, we can choose $c_2=-1/12$ and in order to make the scalar field to have the dimension of mass we can scale the parameters respectively $c_3=\tilde c_3/\Lambda_3^3\dots$ etc and reabsorb the dimensionless parameters $\tilde c_n$. The Galileon field $\phi$ can couple to matter $\chi$ with an arbitrary coupling. In this work we will consider the important representatives of commonly used coupling classes. First of all, a natural way of coupling the Galileon field to the matter is through a conformal coupling of the form $\phi T$ where $T=T^\mu_\mu$ is the trace of the associated stress energy tensor of the matter field $\chi$. As we mentioned, we assume that the fields are propagating on flat Euclidean space-times and therefore indices are raised and lowered by the Euclidean metric $T=\delta^{\mu\nu}T_{\mu\nu}$. The stress energy tensor for our matter field is simply given by
\begin{equation}
T_{\mu\nu}=\frac{-2}{\sqrt{-g}}\frac{\delta \mathcal{L}_{\rm matter}\sqrt{-g} }{\delta g^{\mu\nu}}|_{g=\delta}=-\partial_\mu \chi \partial_\nu \chi +\frac 12 \delta_{\mu\nu} \left((\partial \chi)^2+ M^2 \chi^2\right)
\end{equation}
with its trace being $T=((\partial\chi)^2+2M^2\chi^2)$, such that the conformal coupling reads
\begin{equation}
\frac{\phi((\partial\chi)^2+2M^2\chi^2)}{M_C}
\end{equation}
We suppressed the interaction by a so far arbitrary scale $M_C$, which quotes when this interaction becomes important. In the context of massive gravity, the coupling of the Galileon in massive gravity (the helicity-0 degree of freedom of the massive graviton) with the matter field comes from the coupling $h_{\mu\nu}T^{\mu\nu}/M_{\rm Pl}$, which would mean that the $\phi T$ coupling is Planck mass suppressed $M_C=M_{\rm Pl}$. However, if we consider the Galileon field as a scalar field in its own right, then this scale can be arbitrarily different from the Planck mass. This conformal coupling can be extended to the non-linear level $\phi^2 T$ at the prize of loosing the Galileon symmetry. From the next order stress energy tensor one can also construct this type of non-linear conformal couplings $\phi^2 T^{\mu\;\;\alpha}_{\;\;\;\mu\;\;\;\alpha}$ where the tensor $T^{\mu\nu\alpha\beta}$ arises from the variation of $T_{\mu\nu}$ with respect to the metric,
\begin{eqnarray}
T_{\mu\nu\alpha\beta}&=&\frac{-2}{\sqrt{-g}}\frac{\delta\sqrt{-g}T_{\mu\nu}}{\delta g^{\alpha\beta}}|_{g=\delta}=- \partial_\mu \chi \partial_\nu \chi \delta_{\alpha\beta}- \partial_\alpha \chi \partial_\beta \chi \delta_{\mu\nu}\\ 
&+&\frac12 \left(\delta_{\mu\alpha}\delta_{\nu\beta}+\delta_{\nu\alpha}\delta_{\mu\beta}+\delta_{\mu\nu}\delta_{\alpha\beta}\right)((\partial \chi)^2+ M^2 \chi^2)\,,
\end{eqnarray}
such that the non-linear conformal coupling would become
\begin{equation}
\frac{\phi^2(4(\partial\chi)^2+12M^2\chi^2)}{M^2_{NC}}
\end{equation}
suppressed by the scale $M^2_{NC}$. Again, in the context of massive gravity, this coupling would arise from $h_{\mu\nu}h_{\alpha\beta}T^{\mu\nu\alpha\beta}/M^2_{\rm Pl}$ but here we shall keep the scale arbitrary. Another important class of possible couplings is the derivative coupling. At the linear level we can also have longitudinal coupling of the form $\partial_\mu \partial_\nu \phi T^{\mu\nu}$ which would not contribute in the case of conserved sources, and at the non-linear level we would respectively have the disformal coupling of the form $\partial_\mu \phi \partial_\nu \phi T^{\mu\nu}$. This derivative coupling naturally arises in the massive gravity and  usually comes hand in hand with interesting features. \cite{deRham:2010ik, deRham:2010tw}. In massive gravity, this interaction would come in at a scale $\partial_\mu \phi \partial_\nu \phi T^{\mu\nu}/(M_{\rm Pl}\Lambda_3^3)$, but here we shall consider this coupling at an arbitrary scale $1/M^4_{D}$. 
\begin{equation}
\frac{-\partial^\mu \phi \partial^\nu \phi\partial_\mu \chi \partial_\nu \chi +\frac 12 (\partial \phi)^2\left((\partial \chi)^2+ M^2 \chi^2\right)}{M^4_{D}}
\end{equation}
In the context of Horndeski interactions and screening mechanisms this disformal coupling has also received much attention \cite{Koivisto:2012za, Zumalacarregui:2013pma, Bettoni:2014ana, Brax:2014vva, Burrage:2014uwa}.

For the one loop calculations we will need the Feynman propagators for the Galileon and the scalar field. The Galileon field is a massless scalar field and therefore its Feynman propagator is simply given by
\begin{equation}\label{procGal}
G_{\phi}=\langle \phi(x_1) \phi(x_2)\rangle=\int \frac{\d^4 p}{(2\pi)^4}\frac{e^{i p_\mu \left(x_1^ \mu - x_2^\mu\right)}}{p^2}\,.
\end{equation}
On the other hand the matter field we consider here is a simple massive scalar field with the mass $M$
\begin{eqnarray}
G_{\chi}&=&\langle \chi(x_1) \chi(x_2)\rangle=\int \frac{\d^4 k}{(2\pi)^4}\frac{e^{i k_\mu \left(x_1^ \mu - x_2^\mu\right)}}{k^2+M^2}\,.
\end{eqnarray}

Before starting the computation, let us first remind us of the the following useful general property due symmetry when computing integrals over loops:
\begin{eqnarray}
\frac{1}{M^4} \int\frac{\d^4k}{(2\pi)^4} \frac{k^{2n}k_{a_1}k_{b_1}\cdots k_{a_m}k_{b_m}}{\left(k^2+M^2\right)^{n+m}}=\frac{1}{2^m (m+1)!}\delta_{a_1b_1\cdots a_mb_m} J_{n+m}\,.
\end{eqnarray}
where $J_n$ stands for the notation
\begin{eqnarray}
J_n=\frac{1}{M^{4}}\int \frac{\d^4 k}{(2\pi)^4} \frac{k^{2n}}{\left(k^2+M^2\right)^n}\,.
\end{eqnarray}
The for our purpose specifically useful cases are mostly
\begin{eqnarray}
\label{Juv}
&& \frac{1}{M^{4}}\int \frac{\d^4 k}{(2\pi)^4} \frac{k^{2(n-1)}k_\mu k_\nu}{\left(k^2+M^2\right)^n}=\frac 1 4 \delta_{\mu\nu} J_{n}\\
&& \frac{1}{M^{4}}\int \frac{\d^4 k}{(2\pi)^4} \frac{k^{2(n-2)}k_\mu k_\nu k_\alpha k_\beta }{\left(k^2+M^2\right)^n}=\frac 1{24}
\delta_{\mu\nu\alpha\beta}J_{n}\\
&& \frac{1}{M^{4}}\int \frac{\d^4 k}{(2\pi)^4} \frac{k^{2(n-3)}k_\mu k_\nu k_\alpha k_\beta k_\delta k_\gamma}{\left(k^2+M^2\right)^n}=\frac 1{192} \delta_{\mu\nu\alpha\beta\delta\gamma}J_{n}\,,
\end{eqnarray}
with
\begin{eqnarray}
&&\delta_{\mu\nu\alpha\beta}=\left(\delta_{\mu\nu} \delta_{\alpha\beta}+\delta_{\mu\alpha}\delta_{\nu\beta}+\delta_{\mu\beta}\delta_{\nu \alpha}\right)\\
&&\delta_{\mu\nu\alpha\beta\delta\gamma}=\left(\delta_{\mu\nu} \delta_{\alpha\beta\delta\gamma}+\delta_{\mu\alpha} \delta_{\nu\beta\delta\gamma}+\delta_{\mu\beta} \delta_{\alpha\nu\delta\gamma}+\delta_{\mu\delta} \delta_{\alpha\beta\nu\gamma}+\delta_{\mu\gamma} \delta_{\alpha\beta\delta\nu}\right)\,.
\end{eqnarray}
In this work we will be interested in the running of the interactions such that we will use dimensional regularization (or equivalently focuse on the log contribution from cutoff regularization). In this case we have the relation
\begin{equation}
J_n=\frac{n(n+1)}{2} J_1\,.
\end{equation}
Another very useful formula that we will be using throughout the paper is the following identity
\begin{equation}\label{masterformula}
\frac{1}{A_1^{\alpha_1}A_2^{\alpha_2}\cdots A_n^{\alpha_n}}= \int_0^1 dx_1\cdots dx_n \delta\left(\sum x_i-1\right) \frac{\prod x_i^{\alpha_i-1}}{\left(\sum x_i A_i\right)^{\sum \alpha_i}}  \frac{\Gamma(\alpha_1+\cdots \alpha_n)}{\Gamma(\alpha_1)\cdots\Gamma(\alpha_n)}
\end{equation}
with the Feynman parameters $x_i$.

%

\section{Galileon Loops}
The quantum corrections coming from the Galileon self interactions have already been extensively studied in the literature and it was successfully shown that they are protected under quantum corrections \cite{Nicolis:2004qq, Nicolis:2008in, Hinterbichler:2010xn, deRham:2012ew, Brouzakis:2013lla, dePaulaNetto:2012hm}. First of all, the symmetry of the theory, namely the shift and Galileon symmetry, prevents to generate local operators by loop corrections which breaks explicitly this symmetry. Nevertheless, this does not forbid to generate local operators which are invariant under shift and Galileon transformations, either renormalizing Galileon interactions themselves or generating operators of higher derivative interactions. It is rather the non-renormalization theorem, which ensures that the Galileon interactions themselves are not renormalized and that the higher derivative operators are irrelevant corrections in the regime of validity of the effective field theory $\partial^n\phi\ll\Lambda_3^{n+1}$ for $n\ge3$. One can show the non-renormalization theorem in a straightforward way by realizing that each vertex in an arbitrary Feynman diagram gives rise to interactions with at least one more derivative. \\

Without loss of generality, let us for a moment concentrate on the cubic Galileon interaction $\phi{\mathcal{E}}^{\mu\alpha\rho\sigma}{{\mathcal{E}}^{\nu\beta}}_{\rho\sigma}\Phi_{\mu\nu}\Phi_{\alpha\beta}/\Lambda_3^3$. We can quickly compute the 2-point function contributions coming from this interaction. We will let the field with the two derivatives run in the loop.
\begin{center}
\begin{figure}[h]
\begin{center}
 \includegraphics[width=7cm]{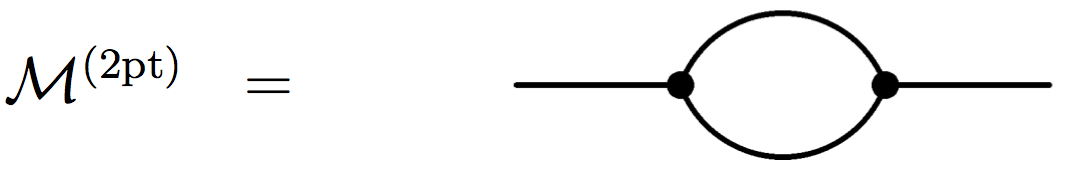}
\end{center}
  \caption{1-loop contribution to the 2-point correlation function from the cubic Galileon interaction. Solid lines denote the Galileon scalar field.}
 \label{Feynman_Gal3}
\end{figure}
\end{center}

\begin{eqnarray}
\mathcal{M}^{\rm (2pt,3vt)}_{\Lambda_3^3} &=& \frac {(-1)^2}{2 !\Lambda_3^6} \left(2 \cdot 2\right)\int  \frac{\d^4 k}{(2\pi)^4} \frac{(4k^\alpha k^\beta q_{1\mu}q_{1\nu}(k_\alpha k_\beta q_1^{\mu}q_1^{\nu}+k^\mu q_{1\beta}(k^\nu q_{1\alpha}-2k_\alpha q_1^\nu)))}{\left(k^2+q_1^2\right)}   \nonumber\\
&=&\frac{2}{\Lambda_3^6}\left(\frac{p_1^8}{8} (3J_1) \right) \,,
\end{eqnarray}
The counter term generated by this diagram contains more derivatives per field $(\partial^4\phi)^2/\Lambda_3^6$ than the classical interaction $\Box\phi(\partial\phi)^2/\Lambda_3^3$. We can generalize this result to an arbitrary Feynman diagram with this interaction at a given vertex. We can contract an external Galileon field $\phi$ with momentum  $p_\mu$ with the Galileon field coming without derivatives in this vertex from the cubic Galileon while letting the other two $\phi$-particles run in the loop with momenta $k_\mu$ and $(p+k)_\mu$. 
The contribution of this vertex to the scattering amplitude is \cite{deRham:2012ew}
\begin{eqnarray}
\mathcal A
 \propto  \int \frac{\mathrm{d}^4 k}{(2\pi)^4} G_k \,
 G_{k+p}\  \, {\mathcal{E}}^{\mu\alpha\rho\sigma}
{{\mathcal{E}}^{\nu\beta}}_{\rho\sigma}
\, \, k_\mu \, k_\nu \, (p+k)_\alpha \, (p+k)_\beta \cdots \,,
\end{eqnarray}
where $G_k$ is the Feynman massless propagator for the Galileon field \ref{procGal}. Now, it is a trivial observation that all the terms which are linear in the external momentum ${\mathcal{E}}^{\mu\alpha\rho\sigma}
{{\mathcal{E}}^{\nu\beta}}_{\rho\sigma}k_\mu k_\nu k_\alpha p_\beta$ as well as all the contributions which are independent of it ${\mathcal{E}}^{\mu\alpha\rho\sigma}
{{\mathcal{E}}^{\nu\beta}}_{\rho\sigma}k_\alpha k_\beta k_\mu k_\nu$ will cancel owing to the antisymmetric nature of the vertex (carried by the indices in the Levi-Civita symbol). Therefore, the only non-vanishing term will come in with at least two powers of the external Galileon field with momentum $p_\alpha p_\beta$. This is the essence of the non-renormalization theorem of the Galileon interactions. In a similar way, if we contract the external leg with the derivative free field in vertices of quartic and quintic Galileon, the same argument straightforwardly leads to the same conclusion regarding the minimal number of derivatives on external fields.

\begin{equation}
\mathcal A
 \propto  \int \frac{\mathrm{d}^4 k}{(2\pi)^4} G_{k_1} \,
 G_{k_2}\cdots\  \, {\mathcal{E}}^{\mu_1\dots \mu_4}
{{\mathcal{E}}^{\nu_1\dots \nu_4}} \prod_{i=1}^{n-1}k_{\mu_i} k_{\nu_i}
\, \, (p+k+\dots)_{\mu_n} \, (p+k+\dots)_{\mu_n} \,, \prod_{j=n+1}^4 \eta_{\mu_j\nu_j}
\end{equation}
 There is no counterterm which takes the Galileon form, and the Galileon interactions are hence not renormalized. This guaranties the stability of the Galileon interactions under quantum corrections if one only considers Galileon loops. However, this is not the case when one starts considering couplings to other matter fields. In the following section we will study in great detail the quantum corrections coming from matter loops.


\section{Matter Loops}
In this section we would like to have a look to the quantum corrections coming from matter one-loops. We will compute explicitly the first order loop corrections coming from the tadpole and two point functions ...etc. and try to generalize the results to the case of n-point functions. In General Relativity the helicity-2 degree of freedom can couple to the matter fields as $h^{\mu\nu}T_{\mu\nu}/M_{\rm Pl}$ and $h^{\mu\nu}h^{\alpha\beta}T_{\mu\nu\alpha\beta}/M_{\rm Pl}^2$... etc. Now, if we have an additional propagating scalar degree of freedom as the Galileon field, it can also couple to the matter fields unless we fine-tune the coupling to be zero. The natural way of coupling this additional scalar field to the matter field is via a conformal coupling $\phi T^{\mu}_{\mu}/M_{C}$ and $\phi^2 T^{\mu\;\;\alpha}_{\;\;\;\mu\;\;\;\alpha}/M_{NC}^2$ $\dots$ etc. If this scalar degree of freedom couples to ordinary matter, then it  can mediate a fifth force with a long range of interaction which has never been detected in Solar System gravity tests or laboratory experiments. On that account, it is crucial to find ways to hide this extra degree of freedom on small scales. One could fine-tune its coupling to matter which is less satisfactory. Fortunately, there exist alternatives to fine-tunings thanks to the screening mechanisms that allow to hide the scalar field on small scales while being unleashed on large scales to produce cosmological effects. Typical examples of screening mechanisms are Vainshtein, chameleon or symmetron \cite{Vainshtein:1972sx, Khoury:2003rn, Hinterbichler:2010es}.  One can basically use the mass term, the coupling to matter or the kinetic term of the scalar field to screen its effect. For the Galileon scalar field it is the Vainshtein mechanism which is at work. Besides these two couplings we will also consider couplings of the form $\frac{1}{M_\star^2}\phi^2 T^{\mu}_{\mu}$ suppressed with a different scale $M_\star$. As we already mentioned Galileon interactions naturally arise in the decoupling limit of massive gravity with a very specific way of coupling to matter. Therefore motivated by massive gravity, we will also consider disformal couplings of the form $\frac{1}{M_{D}^4}\partial_\mu \phi \partial_\nu \phi T^{\mu\nu}$ between the Galileon field and the matter fields. Finally we will also consider longitudinal couplings of the form $\frac{1}{M_L^3}\partial_\mu \partial_\nu \phi T^{\mu\nu}$.
\subsection{Conformal coupling}
\begin{center}
\begin{figure}[h]
\begin{center}
 \includegraphics[width=7cm]{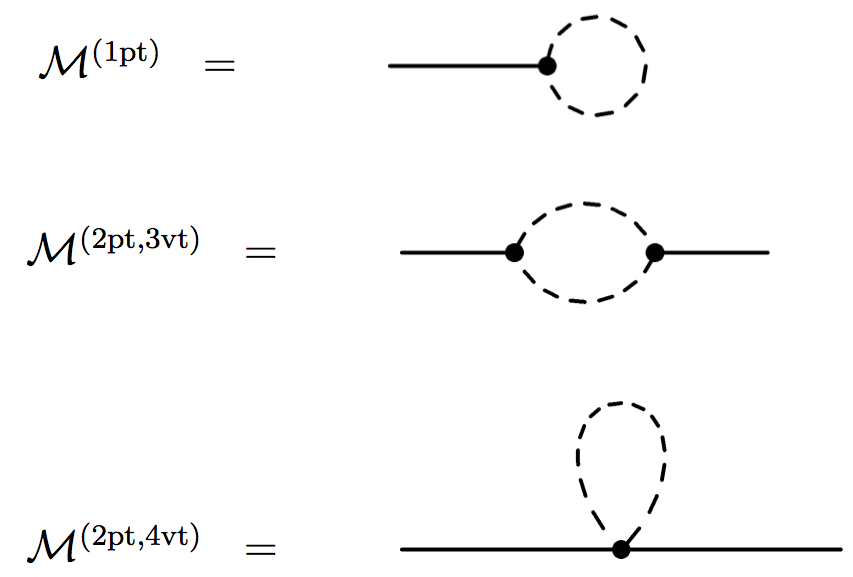}
\end{center}
  \caption{1-loop contributions to the tadpole and 2-point correlation function from the matter coupling. Solid lines denote the Galileon scalar field whereas dashed lines the matter field.}
 \label{Feynman_diagrams}
\end{figure}
\end{center}

Our main interest in this work is the quantum corrections coming from the matter loops. We will consider one loop corrections where only matter field runs in the loop. In this subsection we will first study the quantum corrections coming from conformal couplings. To start let us have a close look at the 1-loop contributions to the tadpole and 2-point correlation function. The corresponding Feynman diagrams are represented in Fig.~\ref{Feynman_diagrams}. We designate by $\mathcal{M}^{\rm (1pt)}$ the 1-loop contribution to the tadpole, by $\mathcal{M}^{\rm (2pt,3vt)}$ the 1-loop correction to the 2-point correlation function arising from the cubic vertex $\phi T/M_C$ and by $\mathcal{M}^{\rm (2pt,4vt)}_{\mu\nu\alpha\beta}$ the 1-loop correction to the 2-point correlation function arising from the quartic vertex $\phi^2 T^{\mu\;\;\alpha}_{\;\;\;\mu\;\;\;\alpha}/M_{NC}^2$ and $\phi^2 T/M_\star^2$.\\
{\bf Zeroth order in external momenta:}\\
To zeroth order in the external momenta we can easily compute the 1-loop contribution to the tadpole from the conformal coupling $\phi T/M_C$ (as represented on the first line of Fig.~\ref{Feynman_diagrams}) yielding
\begin{eqnarray}
\mathcal{M}^{\rm (1pt)}_{p=0, M_{\rm C}} = \frac{ (-1)}{M_{\rm C}}\int \frac{\d^4 k}{(2\pi)^4} \frac{ (k^2+2M^2)}{k^2+M^2} 
=\frac 1 2 \frac{M^4}{M_{\rm C}} J_1 \,.
\end{eqnarray}
Now turning to the 2-point correlation function, the 1-loop contribution  arising for the cubic vertex $\frac{\phi T}{M_{\rm C}}$  to zeroth order in the external momentum is given by
\begin{eqnarray}
\mathcal{M}^{\rm (2pt,3vt)}_{p=0, M_{\rm C}} &=& \frac {(-1)^2}{2 !M_{\rm C}^2} \left(2 \cdot 2\right)\int  \frac{\d^4 k}{(2\pi)^4} \frac{(k^4 +4k^2M^2+4M^4)}{\left(k^2+M^2\right)^2}   \nonumber\\
&=&\frac{2}{M_{\rm C}^2}\left(-\frac{M^4}{2} (3J_1) \right)=-3\frac{M^4}{M_{\rm C}^2}J_1 \,,
\end{eqnarray}
where the terms in bracket in the first line are combinatory factors. Similarly, focusing on the contribution from the quartic vertex, $\frac{\phi^2 T^{\mu\;\;\alpha}_{\;\;\;\mu\;\;\;\alpha}}{M_{\rm NC}^2}$, we obtain
\begin{eqnarray}
\mathcal{M}^{\rm (2pt,4vt)}_{p=0, M_{\rm NC}} &=& \frac {(-1)\left(2\right)}{M_{\rm NC}^2} \int  \frac{\d^4 k}{(2\pi)^4} \frac{\left( 4k^2+12M^2\right) 
}{k^2+M^2}  \nonumber\\
&=&-\frac{2}{M_{\rm NC}^2}\left(-4M^4J_1\right) .
\end{eqnarray}
Finally, from the non-linear conformal coupling $\frac{\phi^2 T}{M_\star^2}$ we have a second contribution to the two point function from the quartic vertex
\begin{eqnarray}
\mathcal{M}^{\rm (2pt,4vt)}_{p=0, M^\star} &=& \frac {(-1)\left(2\right)}{M^\star} \int  \frac{\d^4 k}{(2\pi)^4}  \frac{ (k^2+2M^2)}{k^2+M^2}  \nonumber\\
&=&\frac{M^4}{M_\star^2}J_1 .
\end{eqnarray}
The total 1-loop contribution to the tadpole and 2-point function is thus given by
\begin{eqnarray}
\mathcal{M}^{\rm (1pt)} &=&\frac 1 2 \frac{M^4}{M_{\rm C}} J_1 \\
\mathcal{M}^{\rm (2pt)}&=& \mathcal{M}^{\rm (2pt,3vt)}+ \mathcal{M}^{\rm (2pt,4vt)}  =M^4 \left( \frac{-3}{M_{\rm C}^2}+\frac{8}{M_{\rm NC}^2}+ \frac{1}{M_\star^2}\right)J_1\ \,.
\end{eqnarray}
This corresponds to the following counter-terms at the level of the action
\begin{eqnarray}
\label{LCT}
\L_{CT}&=&-\left(\mathcal{M}^{\rm (1pt)} (\phi)+ \mathcal{M}^{\rm (2pt)} (\phi^2)+\cdots\right)\\
&=&-M^4J_1 \left(\frac{1}{2M_{\rm C}} (\phi)+ \left( \frac{-3}{M_{\rm C}^2}+\frac{8}{M_{\rm NC}^2}+ \frac{1}{M_\star^2}\right)\left(\phi^2\right)+\cdots\right)\\
\label{CC}
\end{eqnarray}
At this point it is not necessary to compute the 3-point or higher n-point functions explicitly. From the tadpole and the 2-point function it is already very suggestive that the coupling $\phi T$ gives rise to appearance of $\phi$, $\phi^2$, $\phi^3$, $\phi^4$ terms..etc, potential interactions for $\phi$ to zeroth order in the external momenta.(The coupling $\phi T$ breaks spontaneously the scale invariance).  It is a trivial observation that the higher n-point functions will give rise to counter terms of higher potential terms 
\begin{eqnarray}
\label{LCTpotential}
\L_{CT}=M^4 J_1\sum_n \left(\frac{1}{M_{\rm C}^n} +\frac{1}{M_{\rm NC}^n} +\frac{1}{M_\star^n}\right)\phi^n
\end{eqnarray}
 These quantum contributions are all suppressed by their own scales $M_{\rm C}$, $M_{\rm NC}$ and $M_\star$ and hence the scale at which they become important is when $M\approx M_{\rm C}^{n/4}$, $M\approx M_{\rm NC}^{n/4}$ or $M\approx M_\star^{n/4}$ respectively. If the scalar field $\phi$ is not an arbitrary scalar, but descends from a full-fledged  tensor field $h_{\mu\nu}$ as in massive gravity, then all these counter terms coming from one-loop matter quantum corrections are Planck mass suppressed
 \begin{equation}
\L^{MG}_{CT}=M^4 J_1\sum_n \frac{1}{M_{\rm Pl}^n} \phi^n
\end{equation}
and so for a Galileon coming from massive gravity one can ignore these quantum corrections as long as the mass of the external matter field is smaller than Planck mass.\\
 {\bf Non-vanishing external momenta:}\\
 Now we would like also to know what kind of operators are generated when we do not put the external momentum to zero. In this case we will not only obtain quantum corrections in form of potential interactions but also derivative interactions. These generated quantum corrections will renormalize the Galileon interactions but also give rise to higher derivative operators with a ghost. From the conformal coupling  $\frac{\phi T}{M_{\rm C}}$ there will not be any contribution with non-zero external momenta to the tadpole since the momentum running in the loop is unaware of the external momentum. From the tadpole there will be only a contribution in form of a potential linear in the field which we already computed above. Therefore, the first contribution comes from the 2-point function.
 \begin{eqnarray}
\mathcal{M}^{\rm (2pt,3vt)}_{p\ne0, M_{\rm C}} &=& \frac {(-1)^2}{2 !M_{\rm C}^2} \left(2 \cdot 2\right)\int  \frac{\d^4 k}{(2\pi)^4} \frac{(4M^4+k_\mu k^\mu(-4M^2+k^\nu(k_\nu-2p_\nu))+k^\mu p_\mu(4M^2+k^\nu p_\nu))}{\left(k^2+M^2\right)((p-k)^2+M^2)}   \nonumber\\
\end{eqnarray}
In order to perform this integration, we will use the specific relation of \ref{masterformula} with two factors in the denominator
\begin{equation}
\frac{1}{A_1^\alpha A_2^\beta}=\int_0^1 d_x \frac{x^{\alpha -1}(1-x)^{\beta-1}}{\left(xA_1+(1-x)A_2 \right)^{\alpha+\beta}} \frac{\Gamma(\alpha+\beta)}{\Gamma(\alpha)\Gamma(\beta)}.
\end{equation} 
For the above integration we have $\alpha=1$, $\beta=1$, $A_1=k^2+M^2$ and $A_2=(p-k)^2+M^2$. The key point is now to perform a change of variable such that the mixing of the two momentas in $(p-k)^2$ disappears. This is indeed achieved by defining $k_\mu=l_\mu-(x-1)p_\mu$. In this way, the propagators in the denominator becomes simply $(l^2+\Delta^2)^2$ where $\Delta=M^2+p^2x(1-x)$ and we can easily perform the integration. After using the relations \ref{Juv} and performing the $x-$integration we obtain the following contribution\\
  \begin{eqnarray}
\mathcal{M}^{\rm (2pt,3vt)}_{p\ne0, M_{\rm C}} &= \frac{2}{M_{\rm C}^2}\frac{(3J_1)}{8}\left( -4M^4-2M^2p^2+p^4) \right) \nonumber\\
  \end{eqnarray}
  The counter terms arising from the two point function would be thus of the form
   \begin{eqnarray}
\L_{CT}\supset -\frac{3J_1}{4M_{\rm C}^2}\left( -4M^4(\phi)^2+2M^2(\partial\phi)^2+(\Box\phi)^2 \right) 
\end{eqnarray}
 The first contribution in form of a potential corresponds to the contribution we computed above to zeroth order in external momentum. Now with a non-vanishing external momentum we obtain derivative interactions. The quantum corrections coming from this 2-point function renormalize the kinetic term of the Galileon field but also generates dangerous higher order derivative interactions $(\Box\phi)^2$. This operator will become important and hence need to be considered at a scale $\partial^2 \sim M_C$.\\
 Similarly as the tadpole, the two point function with the quartic vertex (on the last line of Fig.~\ref{Feynman_diagrams} ) coming from the interactions $\frac{\phi^2 T^{\mu\;\;\alpha}_{\;\;\;\mu\;\;\;\alpha}}{M_{\rm NC}^2}$ and $\frac{\phi^2 T}{M_\star^2}$ will not have any contribution with non-zero external momentum since again the internal momentum in the loop is unaware of the external momentum. At the two point function level the above contribution will be thus the only one. In order to capture better the general behavior of the n-point function contribution, we shall also perform at this stage the contribution of the 3-point function with non-zero external momentum. 
\begin{center}
\begin{figure}[h]
\begin{center}
 \includegraphics[width=10cm]{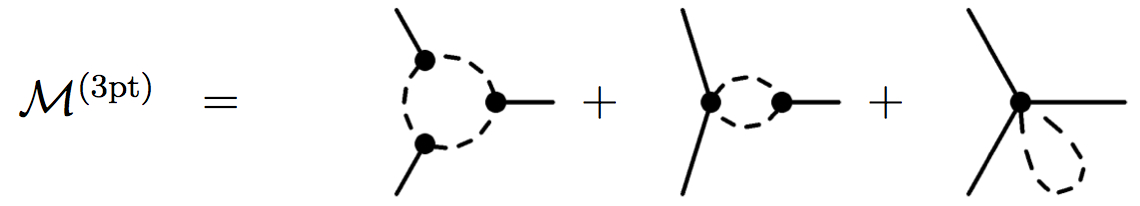}
\end{center}
  \caption{1-loop contributions to the 3-point correlation function from the matter coupling.}
 \label{Feynman_diagrams_3pt}
\end{figure}
\end{center}
In Fig. \ref{Feynman_diagrams_3pt} we see the three different contributions to the 3-point function. The first diagram constitutes three vertices with the conformal coupling $\phi T/M_{\rm C}$ giving rise to a contribution of the following form
 \begin{eqnarray}
\mathcal{M}^{\rm (3pt,3vt)}_{p\ne0} &=& \frac {(-1)^3}{3 !M_{\rm C}^3} \left(3 \cdot 2\right) \left(4 \cdot 2\right)\int  \frac{\d^4 k}{(2\pi)^4} \frac{\mathcal{J}}{\left(k^2+M^2\right)(q_1^2+M^2)(q_2^2+M^2)}   \nonumber\\
\end{eqnarray}
with the momentum conversation relations $q_1=(p_1-k)$ and $q_2=(k+p_2)$ and the shortcut notation
\begin{equation}
\mathcal{J}=8M^6+4M^4q_1^\mu q_{2\mu}+k^\mu(2M^2q_{2\mu}(2M^2+q_1^\nu q_{2\nu})+q_{1\mu}(4M^4+q_{2\nu}(2M^2(k^\nu+q_1^\nu)+k^\nu q_1^\alpha q_{2\alpha})))
\end{equation}
In order to perform this integration involving a denominator with three factors we again use equation  \ref{masterformula} 
\begin{equation}\label{master_formulae_3}
\frac{1}{A_1 A_2 A_3}=\int_0^1 dx_1 dx_2\frac{(3-1)!}{\left(x_1A_1+x_2A_2+(1-x_1-x_2)A_3\right)^3}
\end{equation}
Similarly as we did for the 2-point function, we use the trick of completing the square by shifting the integration variable $k$ to $k_\mu=l_\mu-(p_{2\mu}-p_{2\mu}x_1-p_{1\mu}x_2-p_{2\mu}x_2)$ in order to absorb the terms which are linear in $p_1$ and $p_2$. The denominator then again simply become $(l^2+\Delta^2)^3$ with this time $\Delta=M^2-p_2^2(x_1-1)x_1+(p_1+p_2)(p_1+p_2-2p_2x_1)x_2-(p_1+p_2)^2x_2^2$. We can now perform the integration over the momentum $l_\mu$ and the Feynman parameters $x_1$ and $x_2$. This results in
 \begin{eqnarray}
\mathcal{M}^{\rm (3pt,3vt)}_{p\ne0} &=&  \frac {-(6J_1)}{6M_{\rm C}^3}\left(192M^4+3p_1^4+4p_1^2p_2^2+132p_2^4+18M^2(p_1^2-11p_2^2)\right. \nonumber\\
&&\left. +2p_1\cdot p_2(3(-41M^2+p_1^2+44p_2^2)+67p_1\cdot p_2) \right)
\end{eqnarray}
In an analog way, to the second diagram in fig. \ref{Feynman_diagrams_3pt} the interactions $\frac{\phi^2 T^{\mu\;\;\alpha}_{\;\;\;\mu\;\;\;\alpha}}{M_{\rm NC}^2}$ and $\phi T/M_{\rm C}$ will contribute at the respective vertex yielding
 \begin{eqnarray}
\mathcal{M}^{\rm (3pt,4vt-3vt)}_{p\ne0} &=& \frac {(-1)^2}{2 !M_{\rm NC}^2M_{\rm C}} \left(3 \right) \left(2\right)\int  \frac{\d^4 k}{(2\pi)^4} \frac{4\left(6M^4+k^\mu(p_{1\mu}+p_{2\mu}-k_\mu)(5M^2+k^\nu(p_{1\mu}+p_{2\mu}-k_\nu))\right)}{\left(k^2+M^2\right)((p_1+p_2-k)^2+M^2)}   \nonumber\\
&=&\frac{3J_1}{2M_{\rm NC}^2M_{\rm C}}\left( 228 M^4+147p_1^4-58p_1^2p_2^2+3p_2^4+48M^2(-10p_1^2+p_2^2) \right. \nonumber\\
&& \left. +4p_1\cdot p_2(3(8M^2-7p_1^2+p_2^2)+7p_1\cdot p_2)\right)
\end{eqnarray}
From the 2- and 3- point function one clearly sees that the one loop contributions are always proportional to $M^4$, $M^2 \partial^2$ or $\partial^4$. The contributions proportional to $M^4$ we can ignore here at this stage since they correspond to the contributions from \ref{LCTpotential} in the zero external momentum limit. On the other hand the contributions with the scaling $M^2\partial^2$ will be harmless in the sense that they will give rise to few derivatives per field. The dangerous terms are hence the terms with higher derivatives per field. Thus, the coupling to matter will give rise to counter terms with the following dangerous higher order derivatives acting on the Galileon field.
 \begin{eqnarray}
\label{LCTp}
\L_{CT}&=&-\left( \mathcal{M}^{\rm (2pt)} (\phi^2)+ \mathcal{M}^{\rm (3pt)} (\phi^3)\cdots\right)  \nonumber\\
&&\supset M^2\frac{(\partial \phi)^2}{M_{\rm C}^2}+\frac{(\partial^2 \phi)^2}{M_{\rm C}^2}+ \phi\frac{(\partial^2 \phi)^2}{M_{\rm C}^3}+ \frac{\Box\phi(\partial \phi)^2}{M_{\rm NC}^2M_{\rm C}}+ \phi\frac{(\partial^2 \phi)^2}{M_{\rm NC}^2M_{\rm C}}+\phi\frac{(\partial_\mu\partial_\nu \phi)^2}{M_{\rm C}^3} \cdots
\end{eqnarray}
The quantum corrections coming from the 2-point function renormalizes the kinetic term with a scale $M^2/M_C^2$, whereas the 3-point functions renormalizes the cubic Galileon interaction with a scale $1/M_C^3$ and $1/M_{\rm NC}^2M_{\rm C}$ respectively. Therefore, the scale of the interactions need to be chosen above the scale $\Lambda_3$ below which the Galileon interactions are important. For scales $M_{\rm C}\approx \Lambda_3$, the nice structure of the Galileon interaction would be detuned at an unacceptable scale. For the restricted Galileon field coming from massive gravity, the scale at which this ghost would become accessible is close to the Planck scale. The higher n-point functions would give rise to counter terms of the form
\begin{eqnarray}
\label{LCTpotential}
\L_{CT}=J_1\sum_n \left(\frac{\partial^2\phi}{M_{\rm Pl}} \right)^n
\end{eqnarray}
Near sources these quantum corrections are even more suppressed due to the Vainshtein mechanism $(\Box\phi)^{n-2}/(M_{\rm Pl}^nZ)$.
In massive gravity this is completely harmless since all these quantum corrections will start playing a role close to the Planck mass $M_{\rm Pl}$.\\

\subsection{Disformal coupling}
We would like now to draw our attention to the derivative coupling of the form $\partial_\mu \phi \partial_\nu \phi T^{\mu\nu}/M_D^4$. This coupling has to be considered in the context of massive gravity. There it arises naturally after shifting the metric perturbations by $h_{\mu\nu} \to \phi\eta_{\mu\nu}+\partial_\mu \phi \partial_\nu \phi$ in order to diagonalize some of the interactions between the helicity-2 and helicity-0 degrees of freedom. In massive gravity this derivative disformal coupling brings along important consequences. It plays a very crucial role for the existence of degravitating solutions \cite{deRham:2010tw} which is absent in the usual Galileon theories. Furthermore, it plays an important role in the lensing measurements since photons can now couple to the scalar degree of freedom \cite{2011PhRvL.106t1102W}.\\
We would like now to study the quantum corrections coming from this derivative coupling. To zeroth order in the external momentum this coupling will give zero contributions and hence the potential terms from \ref{LCTpotential} will not be generated. Also since already two $\phi$ fields appear in the coupling, there will not be any contribution to the tadpole either, or to any odd n-point function. Thus, there will be only contributions to the even number n-point functions. The first contribution will be to the 2-point function tadpole depicted in the last line of fig. \ref{Feynman_diagrams}.  
\begin{eqnarray}
\mathcal{M}^{\rm (2pt,4vt)}_{D} = \frac{(-1)^1(2)p_\mu p_\nu}{M_{\rm D}^4} \int  \frac{\d^4 k}{(2\pi)^4} \frac{\left(-k^\mu k^\nu +\frac12\eta^{\mu\nu}(k^2+M^2)\right)}{k^2+M^2}= \frac{M^4}{4M_D^4} p^2
\ \,.
\end{eqnarray}
which would again simply renormalize the kinetic term of the Galileon scalar field $\phi$. As next, let us have a look at the next leading contribution coming from the 4-point function.
\begin{center}
\begin{figure}[h]
\begin{center}
 \includegraphics[width=7cm]{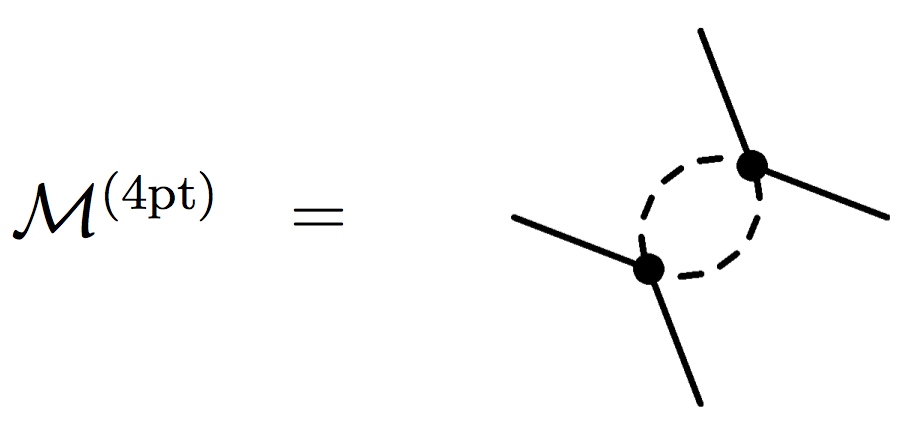}
\end{center}
  \caption{One-loop contributions to the 4-point function coming from the derivative coupling.}
 \label{Feynman_diagrams_4pt}
\end{figure}
\end{center}
\begin{eqnarray}
\mathcal{M}^{\rm (4pt,4vt)}_{D} =\frac{(-1)^2(4\cdot2\cdot2)p^{1\mu} p^{2\nu}p^{3\alpha}p^{4\beta}}{2!M_{\rm D}^8} \int  \frac{\d^4 k}{(2\pi)^4}\frac14 \frac{\mathcal{H} }{(k^2+M^2)(M^2+(p_1+p_2-k)^2)}
\end{eqnarray}
where the shortcut $\mathcal{H}$ stands for the following tensor
 \begin{eqnarray}
 \label{defH}
\mathcal{H}&=&-2k_\mu q_{1\nu}(-2k_\alpha q_{1\beta}+g_{\alpha\beta}(M^2+k\cdot q_1))+g_{\mu\nu}(-2k_{\alpha}q_{1\beta}(M^2+k\cdot q_1)  \nonumber\\
&&+g_{\alpha\beta}(M^4+k\cdot q_1(2M^2+k\cdot q_1)))
\end{eqnarray}
The energy conservation forces $q_1=p_1+p_2-k$ and $p_4=p_1+p_2-p_3$. After performing the integral we obtain the long expression
\begin{eqnarray}
\mathcal{M}^{\rm (4pt)}_{D} =\frac{1}{60M_{\rm D}^8}\left( -2p_1^2 p_2^2(10M^2+3(p_1^2+p_2^2))p_3\cdot p_4-2p_1^2(10M^2+p_1^2-3p_2^2)(p_2\cdot p_3) p_2\cdot p_4  \right.\nonumber\\
+p_1^\alpha(p_2^\beta((30M^4-5p_1^4-2p_1^2 p_2^2-5p_2^4-30M^2(p_1^2+p_2^2))p_{3\alpha}p_{4\beta} \nonumber\\
+p_{3\beta}((30M^4+5p_1^4+18p_1^2 p_2^2 +5 p_2^4+30M^2(p_1^2+p_2^2))p_{4\alpha}-4(10M^2+p_1^2)p_{2\alpha}p_2\cdot p_4)) \nonumber\\
+2((105M^4-6p_1^2 p_2^2+20M^2(p_1^2+p_2^2))p_{2\alpha}p_3\cdot p_4+p_{1}^\beta(-p_2^2(10M^2-3p_1^2+p_2^2)p_{3\alpha}p_{4\beta}  \nonumber\\
+p_{2\alpha}((-2(10M^2+p_2^2)p^\gamma_1 p_{3\beta}+(50M^2+3(p_1^2+p_2^2))p_{2\beta}p_3^\gamma+8p_1^\gamma p_{2\beta}p_2\cdot p_3)p_{4\gamma}  \nonumber\\
-2p_1^\gamma p_{2\beta}(2(p_1^\delta+3p_2^\delta)p_{3\gamma}-3p_{2\gamma}p_3^\delta)p_{4\delta}+p_2^\gamma(-(40M^2+9(p_1^2+p_2^2))p_{3\beta}p_{4\gamma} \nonumber\\
+\left. p_{3\gamma}((20M^2+11(p_1^2+p_2^2))p_{4\beta}-4p_{2\beta}p_2\cdot p_4))))))\right) \nonumber\\
\end{eqnarray}
Even if the above contribution is a very long expression we can still extract useful information. The derivative disformal coupling gives counter terms of the following form
 \begin{equation}
\L_{CT}=-\left(\mathcal{M}^{\rm (2pt)}_{D}(\phi^2)+\mathcal{M}^{\rm (4pt)}_{D} (\phi^4)+\cdots\right) \supset \frac{M^4}{M_{\rm D}^4}(\partial\phi)^2 + \frac{M^2}{M_{\rm D}^8}(\Box\phi)^2 (\partial\phi)^2 \frac{1}{M_{\rm D}^8}\Box\phi \partial^4\phi (\partial\phi)^2+  \cdots
\end{equation}
We see immediately, that the kinetic term and the quartic Galileon interactions get renormalized. Besides that, the quantum corrections generate counter terms with too many derivatives acting per field which do not belong to the Galileon class of interactions and contain a ghost degree of freedom. The kinetic term receives a quantum correction scaling as $M^4/M_D^4$ which becomes important for $M\sim M_D$. The quartic Galileon on the other hand gets renormalized by an operator which scales as $M^2/M_D^8$. The higher derivative operator remains negligible as long as $\partial^4/M_D^8<<\Lambda_3^4$. Again in the case of massive gravity these contributions would be harmless since they would be Planck mass suppressed instead of the $M_D$ suppressed. 
\subsection{Longitudinal coupling}
Even if longitudinal couplings of the form $\frac{1}{M_L^3}\partial_\mu \partial_\nu \phi T^{\mu\nu}$ would vanish for conserved sources and do not originate from theories like massive gravity, it is worth studying quickly the counter terms arising from this type of couplings. The scalar field comes already with two derivatives in this longitudinal coupling and we expect that the counter terms will result in operators with higher derivatives per field. The easiest contribution to compute is the tadpole contribution
\begin{eqnarray}
\mathcal{M}^{\rm (1pt)} &=& \frac{ (-1)}{M_L^3}p_\mu p_\nu\int \frac{\d^4 k}{(2\pi)^4} \frac{\left(-k^\mu k^\nu +\frac12\eta^{\mu\nu}(k^2+M^2)\right)}{k^2+M^2} \nonumber\\
&=& \frac{M^4}{8M_L^3} p^2
\end{eqnarray}
Thus, the tadpole already generates a term of the form $ (\Box\phi)M^4/ M_L^3$. The contribution to the 2-point function can also be performed easily
 \begin{eqnarray}
\mathcal{M}^{\rm (2pt,3vt)}_{p\ne0} &=& \frac {(-1)^2}{2 !M_L^6} \left(2 \cdot 2\right)p^\mu p^\nu p^{2\alpha} p^{2\beta}\int  \frac{\d^4 k}{(2\pi)^4}\frac14 \frac{\mathcal{H}}{\left(k^2+M^2\right)(q_1^2+M^2)}   \nonumber\\
\end{eqnarray}
where $\mathcal{H}$ is given by \ref{defH} and where $q_1=(k-p)$ and $p_2=p$ because of momentum conversation. After performing the integration this gives
 \begin{eqnarray}
\mathcal{M}^{\rm (2pt,3vt)}_{p\ne0} &=& \frac {2}{M_L^6}(3J_1)\frac{9M^4}{16}p^4
\end{eqnarray}
The longitudinal coupling gives rise to quantum corrections with the following counter terms
 \begin{equation}
\label{LCTstab}
\L_{CT}=-\left(\mathcal{M}^{\rm (1pt,3vt)}_{p\ne0}(\phi)+\mathcal{M}^{\rm (2pt,3vt)}_{p\ne0} (\phi^2)+\cdots\right) \supset \frac{M^4}{M_{\rm L}^3}\Box\phi +  \frac{M^4}{M_{\rm L}^6}(\Box\phi)^2+  \cdots
\end{equation}
As you can see, this longitudinal coupling naturally generates higher derivatives acting per scalar field becoming important at a scale close to $M\sim M_L$.
\subsection{Mixed couplings}
We can also quickly study an example of Feynman diagrams in which both the conformal coupling and the derivative couplings contribute at the same time. These diagrams mix the scales of the two couplings, therefore the counter terms of the n-point function will be suppressed by powers of $(M_{\rm D}^4M_{\rm C})^n$ and $ (M_{\rm L}^3M_{\rm C})^n$ respectively. For our purpose, it will be enough to study the first contributions of these mixed Feynman diagrams. Let us first start with the 3-point function contribution coming from the second diagram depicted in fig. \ref{Feynman_diagrams_3pt} where at the one vertex the coupling $\phi T/M_{\rm C}$ and the other vertex the coupling  $\partial_\mu \phi \partial_\nu \phi T^{\mu\nu}/M_{\rm D}^4$ have to be considered. 
\begin{equation}
\mathcal{M}^{\rm (3pt)} =\frac{(-1)^2(3\cdot2)(2\cdot2)p_{1}^\mu p_{2}^\nu}{2!M_{\rm NL}^4M_{\rm C}} \int  \frac{\d^4 k}{(2\pi)^4}\frac12 \frac{\left(-2k_\mu q_{1\nu}(2M^2+k\cdot q_1)+g_{\mu\nu}(2M^4+k\cdot q_1(3M^2+k\cdot q_1))\right)}{(k^2+M^2)(M^2+(p_1+p_2-k)^2)}
\end{equation}
We can very easily perform this integral. This diagram gives the following contribution
\begin{eqnarray}
\mathcal{M}^{\rm (3pt)} &=&\frac{12}{M_{\rm NL}^4M_{\rm C}}\frac{1}{24}\left(-p_1^2p_2^2(6M^2+p_1^2+p_2^2) +p_1\cdot p_2(45M^4-2p_1^2p_2^2+6M^2(p_1^2+p_2^2) \right. \nonumber\\
&& \left. + p_1\cdot p_2(18M^2+p_1^2+p_2^2+2p_1\cdot p_2))\right)
\end{eqnarray}
Similarly, the contribution to the 2-point function with the one vertex being the coupling $\partial_\mu \partial_\nu \phi T^{\mu\nu}/M_{\rm L}^3$ and the other vertex being the the coupling $\phi T/M_{\rm C}$ gives
\begin{equation}
\mathcal{M}^{\rm (2pt)} =\frac{(-1)^2(2)(2\cdot2)}{2!M_{\rm L}^3M_{\rm C}}\frac{M^2}{8} p^2(15M^2 + 2 p^2)
\end{equation}
Summarizing, the Feynman diagrams with the vertices of mixed couplings to the matter field generate in a very similar way contributions as we were obtaining from the purely conformal or disformal coupling diagrams
 \begin{equation}
\L_{CT}=-\left( \mathcal{M}^{\rm (2pt)} (\phi^2)+ \mathcal{M}^{\rm (3pt)} (\phi^3)\cdots\right) \supset \frac{M^4(\partial \phi)^2}{M_{\rm L}^3M_{\rm C}}+\frac{M^2(\Box \phi)^2}{M_{\rm L}^3M_{\rm C}} +\frac{M^2(\partial \phi)^2\Box\phi}{M_{\rm D}^4M_{\rm C}} +\frac{(\partial \phi)^2(\Box\phi)^2}{M_{\rm D}^4M_{\rm C}} + \cdots
\end{equation}
It is a trivial observation that the contributions to the 3-point function take the form of Galileon interactions as well as new derivatives interactions not embedded in the Galileon interactions as we were obtaining above with the difference that this time the kinetic term scales as $M^4/(M_{\rm L}^3M_{\rm C})$, the cubic Galileon as $M^2/(M_{\rm D}^4M_{\rm C})$ and the higher order derivative operators as $1/M_{\rm D}^4M_{\rm C}$.
\section{Stability of the couplings}
\begin{center}
\begin{figure}[h]
\begin{center}
 \includegraphics[width=8cm]{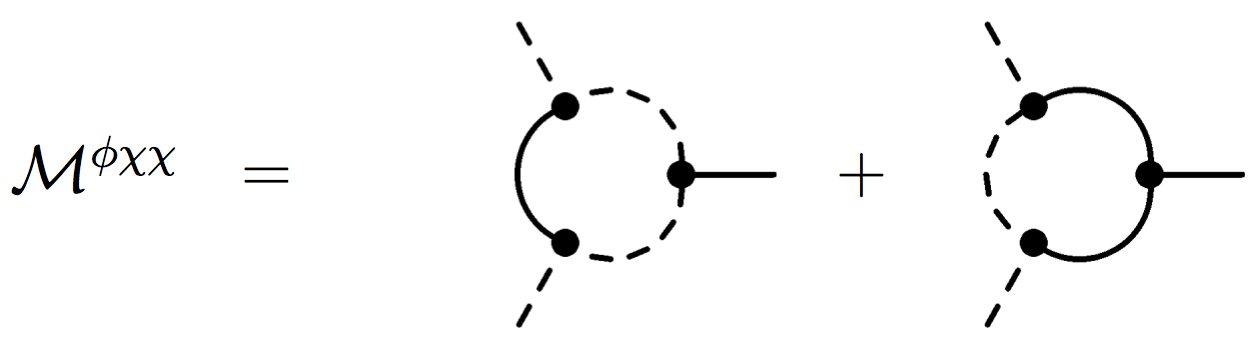}
\end{center}
  \caption{One-loop contributions to the 3-point function renormalizing the coupling itself.}
 \label{Feynman_diagrams_couplStab}
\end{figure}
\end{center}

Another very interesting question is the stability of the classical couplings. If the couplings themselves receive large quantum corrections, it will consequently affect the results we presented above. In order to illustrate the impact of the quantum corrections of the couplings themselves, let us focus on the conformal coupling $\frac{\phi T}{M_{\rm C}}$. The question we want to quickly address is whether or not this coupling receives large quantum corrections. In order to study this question we will consider two specific diagrams with the coupling $\frac{\phi T}{M_{\rm C}}$ at the vertices and also the third Galileon interactions.  The for us interesting diagrams are depicted in Fig. \ref{Feynman_diagrams_couplStab}, where in the first diagram the conformal coupling $\frac{\phi T}{M_{\rm C}}$ acts on the vertices, whereas in the second diagram the third Galileon interaction acts on the second vertex.
The first diagram contributes to  $\langle \phi \chi \chi \rangle$ through the following integration
 \begin{eqnarray}
\mathcal{M}^{ \phi \chi \chi}_1 &=& \frac {(-1)^3}{3 !M_{\rm C}^3}(3)(4\cdot2\cdot2) \int  \frac{\d^4 k}{(2\pi)^4} \frac{  \mathcal{W}}{(k^2+M^2)(q_1^2+M^2)(q_2^2)}   \nonumber\\
\end{eqnarray}
where $\mathcal{W}$ stands for
\begin{equation}
\mathcal{W}=8M^6+4M^4p_3^\mu q_{1\mu}+k^\mu(2M^2q_{1\mu}(2M^2+p_3^\nu q_{1\nu}+p_{2\mu}(4M^4+q_{1\nu}(2M^2(k^\nu+p_3^\nu)+k^\nu p_3^\alpha q_{1\alpha})))
\end{equation}
The conservation of energy dictates $q_{1\mu}=p_{1\mu}-k_\mu$, $q_{2\mu}=p_{2\mu}+k_{\mu}$ and $p_{3\mu}=p_{1\mu}+p_{2\mu}$. Note that now in the propagator with the momentum $q_2$ the mass does not appear since it corresponds to the $\phi$ field running in the loop. Therefore, when we use the trick \ref{master_formulae_3} in order to write the  denominator as $(l^2+\Delta^2)^3$, an important difference will be also in the $\Delta$
\begin{equation}
\Delta^2=M^2(x+x_2)-x_2^2(p_1+p_2)^2+(p_1+p_2)(p_1+p_2-2p_2x)x_2-p_2^2x(-1+x)
\end{equation}
 We can now perform the integration in $l$ and the Feynman parameters easily and find
 \begin{equation}
\mathcal{M}^{ \phi \chi \chi}_1 = \frac {(-8)}{M_{\rm C}^3}\frac{(3J_1)}{8}\left(-16M^4+2p_2^4-M^2(4p_1^2+5p_2^2)+p_1\cdot p_2(-13M^2-p_1^2+4p_2^2+p_1\cdot p_2)  \right)
\end{equation}
In a similar way, we can compute the second diagram in Fig. \ref{Feynman_diagrams_couplStab}. At the two vertices the conformal coupling $\phi T/M_{\rm C}$ acts, whereas on the third vertex the cubic Galileon interaction $\phi \mathcal{E}^{\mu\nu\alpha\beta}\mathcal{E}^{\rho\sigma}_{\;\;\; \alpha\beta}\Phi_{\mu\rho}\Phi_{\nu\sigma}/\Lambda_3^3$. The contribution of this diagram is given by
\begin{eqnarray}
\mathcal{M}^{ \phi \chi \chi}_2 &=& \frac {(-1)^3}{3 !M_{\rm C}^2\Lambda_3^3}(5)(2)(4\cdot2\cdot2) \int  \frac{\d^4 k}{(2\pi)^4} \frac{  \mathcal{W}_2}{(k^2)(q_1^2)(q_2^2+M^2)}   \nonumber\\
\end{eqnarray}
with this time
\begin{eqnarray}
\mathcal{W}_2&=&2k^\mu(k_\mu(4M^4q_{1\nu}q_1^\nu+q_{2\nu}(2M^2(p_2^\nu+p_3^\nu)q_{1\alpha}q_1^\alpha+p_2^\nu p_3^\alpha q_{1\beta}q_1^\beta q_{2\alpha})) \nonumber\\
&&-k^\nu q_{1\mu}q_{1\nu}(4M^4+q_{2\alpha}(2M^2(p_2^\alpha+p_3^\alpha)+p_2^\alpha p_3^\beta q_{2\beta})))
\end{eqnarray}
Two $\phi$-fields are running in the loops, therefore we have two massless propagators with momenta $k_\mu$ and $q_{1\mu}$. We again complete the square by shifting the integration variable $k$ to $k_\mu=l_\mu-(p_{2\mu}-p_{2\mu}x_1-p_{1\mu}x_2-p_{2\mu}x_2)$ and so absorb the linear terms in $p_1$ and $p_2$. The denominator again takes the desired form $(l^2+\Delta^2)^3$ with this time $\Delta=-p_2^2(-1+x_1)x_1+(p_1+p_2)(p_1+p_2-2p_2x_1)x_2-(p_1+p_2)^2x_2^2-M^2(-1+x_1+x_2)$. After performing the integration over the momentum $l_\mu$ and the Feynman parameters $x_1$ and $x_2$ we end up with
\begin{eqnarray}
\mathcal{M}^{ \phi \chi \chi}_2 &=& \frac {(-80)}{3M_{\rm C}^2\Lambda_3^3}\frac{(3J_1)}{16}\left(p_1^2(48M^4+6p_1^2p_2^2+19p_2^4+12M^2(p_1^2+4p_2^2)) \right. \nonumber\\
&& \left. +p_1\cdot p_2(p_1^2(48M^2+3p_1^2+34p_2^2)+4p_1\cdot p_2(3p_1^2-p_2^2-p_1\cdot p_2))  \right)
\end{eqnarray}
From these two diagrams we immediately observe that the conformal coupling itself receives quantum correction in terms of the following counter terms 
 \begin{eqnarray}
\label{LCTstab}
\L_{CT}&=&-\left( \mathcal{M}^{ \phi \chi \chi}_{1} (\phi\chi^2)+ \mathcal{M}^{ \phi \chi \chi}_{ 2}  (\phi\chi^2)+\cdots\right)  \nonumber\\
&&\supset \frac{M^4}{M_{\rm C}^3}\phi \chi^2+  \frac{M^2}{M_{\rm C}^3}\phi \partial_\mu\chi \partial^\mu\chi+  \frac{1}{M_{\rm C}^3}\phi \partial_\mu\partial_\nu\chi \partial^\mu\partial^\nu\chi+   \frac{M^4}{M_{\rm C}^2\Lambda_3^3}\phi \partial_\mu\chi \partial^\mu\chi+  \frac{M^2}{M_{\rm C}^2\Lambda_3^3}(\Box\phi) \partial_\mu\chi \partial^\mu\chi+ \cdots \nonumber\\
\end{eqnarray}
As you can see, the conformal coupling $\phi T/M_{\rm C}$ receives quantum corrections that scale as $M^4/M_{\rm C}^3$ for the mass term and $M^2/M_{\rm C}^3$ for the kinetic term respectively. For masses of the scalar field $\chi$ close to $M\sim M_C$ these quantum corrections to the conformal coupling become very important. The quantum corrections give also rise to couplings with more derivatives of the form $\phi\partial_\mu\partial_\nu\chi \partial^\mu\partial^\nu\chi$. 
\section{Summary and discussion}
Historically, Galileon interactions were discovered by the attempt of generalizing the interactions of the decoupling limit of DGP model \cite{Nicolis:2008in}. They were constructed by insisting on the symmetry of the helicity-0 mode $\phi$ of the DGP model, namely the invariance under internal Galilean and shift transformations. From the perspectives of the higher dimensional induced gravity braneworld models these symmetries can be regarded as residuals of the 5-dimensional Poincar\'e invariance. The invariance under these transformations together with the postulate of ghost-absence restrict the construction of the effective $\phi$ Lagrangian. One can construct only five derivative interactions which fulfill these conditions as descendants of the Lovelock invariants in the bulk of generalized braneworld models \cite{deRham:2010eu, Burrage:2011bt, Goon:2011qf}. The Galileon interactions share a very important property: they do not get renormalized by the quantum corrections arising from the Galileon interactions themselves. It is a common misconception in the literature that the non-renormalization theorem is due to the symmetry of the theory. The symmetry guarantees only that there will not be any quantum corrections in form of potential interactions but the symmetry does not prohibit the generation of the Galileon interactions by the quantum corrections. Maybe it would be more fair to say that the fact that the symmetry is not realized exactly, meaning that the symmetry is fulfilled only up to total derivatives, plays a crucial role for the non-renormalization theorem.  \\

Nevertheless, the non-renormalization theorem does not survive when one includes couplings to matter. One can couple the Galileon scalar field in three different ways to the matter fields: through linear or non-linear conformal coupling, through disformal coupling and finally through a longitudinal coupling. We have considered these three possible ways of coupling the Galileon to the matter field and computed the one-loop quantum corrections to study the counter terms arising from these couplings. Starting with the conformal coupling $\phi T$, we show how the one-loop quantum contributions give rise to potential interactions proportional to the mass of the matter field and suppressed by the scale of the coupling to zeroth order in the external momenta $M^4 \phi^n/M_C^n$. Including the contributions with non-vanishing external momentum we observe how the dangerous operators with higher operators per field are generated $(\Box\phi)^2/M_C^2$. Another important class of couplings is the disformal coupling $\partial_\mu \phi \partial_\nu\phi T^{\mu\nu}$ which arises naturally in the context of massive gravity. Similarly as in the case of conformal couplings, the quantum corrections from this disformal coupling gives rise to higher derivative operators. The quantum corrections coming from a longitudinal coupling $\partial_\mu \partial_\nu\phi T^{\mu\nu}$ share the same destiny. Furthermore, we studied the interesting question whether or not the covariant couplings themselves are stable under quantum corrections. As an example we performed the one-loop quantum corrections to the conformal coupling and found that they get renormalized significantly for scalar masses close to $M\sim M_C^{3/4}$ and $M\sim M_C^{3/2}$ respectively. Additionally, the appearance of new couplings with more derivatives $\phi\partial_\mu\partial_\nu\chi \partial^\mu\partial^\nu\chi$ is unavoidable. \\

Summarizing, in this paper we have shown that quantum corrections coming from the coupling to the matter fields
\begin{enumerate}
\item{\bf renormalize the Galileon interactions:} the non-renoramalization theorem protecting the Galileon interactions does not persist once one considers couplings to matter field. The standard covariant couplings give rise to quantum corrections with non-vanishing external momentum which renormalize the Galileon interactions. 
 \item{\bf generate new higher derivative interactions:} in the same way as the Galileon interactions the coupling to matter fields generates ghost-like interactions with higher derivative operators.
\end{enumerate}
However, within the regime of the effective field theory these renormalized Galileons as well as the new higher derivative interactions come in suppressed by the coupling scale. They are harmless as long as 
\begin{enumerate}
\item the mass of the external scalar field is smaller than the original classical coupling scale $M\ll M_{\rm C}$
 \item{the derivatives applied on the Galileon field is smaller than the original classical coupling scale $\partial \ll M_{\rm C}$}
\end{enumerate}
In the context of massive gravity the original coupling scale is given by the Planck mass and therefore as long as the mass of the scalar field and the derivatives are smaller than the Planck mass, the quantum corrections are totally insignificant. 



\acknowledgments

We would like to thank Claudia de Rham for useful discussions. This work is supported by the Swiss National Science Foundation.


\bibliographystyle{JHEPmodplain}
\bibliography{references}

\end{document}